\documentstyle[prl,aps,floats,twocolumn]{revtex}
\input{psfig.tex}
\begin{document}

\draft
\twocolumn[\hsize\textwidth\columnwidth\hsize\csname @twocolumnfalse\endcsname

\title{Direct Signature of Evolving Gravitational Potential from 
Cosmic Microwave Background }
\author{Uro\v s Seljak\cite{urosemail}\cite{presentaddress}}
\address{
Max-Planck-Institut f\"ur Astrophysik, D--85740 Garching,
Germany }
\author{Matias Zaldarriaga\cite{matiasemail}}
\address{Institute for Advanced Studies, School of Natural Sciences, Princeton, NJ 08540}
\date{September 1998}
\maketitle

\begin{abstract}
We show that time dependent gravitational potential can be directly
detected from the cosmic microwave background (CMB) 
anisotropies. The signature can be measured by cross-correlating the 
CMB with the projected density field reconstructed from the 
weak lensing distortions of the CMB itself. 
The cross-correlation 
gives a signal whenever there is a time dependent gravitational potential. This 
method traces dark matter directly and has a well defined redshift distribution 
of the window projecting over the density perturbations, thereby avoiding the
problems plaguing other proposed cross-correlations.
We show that both MAP and Planck will be able to probe this effect for 
observationally relevant curvature and cosmological constant models, which will provide 
additional constraints on the cosmological parameters.

\end{abstract}
\pacs{PACS numbers: 98.80.Es,95.85.Nv,98.35.Ce,98.70.Vc  \hfill}
]
\def\bi#1{\bbox{#1}}
\def\gsim{\raise2.90pt\hbox{$\scriptstyle
>$} \hspace{-6.4pt}
\lower.5pt\hbox{$\scriptscriptstyle
\sim$}\; }
\def\lsim{\raise2.90pt\hbox{$\scriptstyle
<$} \hspace{-6pt}\lower.5pt\hbox{$\scriptscriptstyle\sim$}\; }

It is widely accepted that 
cosmic microwave background (CMB) anisotropies offer a
unique environment to study cosmological models. The 
anisotropies were generated predominantly during recombination 
at redshift $z \sim 1100$, when the universe was still in 
a linear regime and the physics at eV energy scale was simple.
This allows one to make robust predictions for various 
cosmological models, which can be compared to an increasing 
number of observations. However, some degeneracies between cosmological 
parameters remain even for 
future satellite missions and these are being further expanded as 
new parameters are being introduced. The degeneracies are 
particularly severe between various components affecting 
the expansion of the universe, such as curvature, cosmological 
constant or any other term with a more general equation of state \cite{steinh}.
Other cosmological tests must therefore be used to break these 
degeneracies.

It has been pointed out that in a 
universe where matter density does not equal critical density 
the 
gravitational potential is changing with time,
which produces a 
significant component to the CMB on large scales \cite{kof}. 
This effect is generated at late times and since the gravitational 
potential is related to the density field through the Poisson's 
equation, the effect can also be looked for by 
cross-correlating CMB with another tracer of density field \cite{ct}.
Unfortunately, no clean density map out to high redshift 
exists on large scales. The X-ray background has been suggested as a
possible tracer of large scale structure out to $z\sim 3-4$, but the 
uncertainties associated with the redshift distribution of the sources,
relation between X-ray light and underlying mass
and the Poisson fluctuations from the nearby sources make this 
test inconclusive \cite{bct,kink}. 

Recently we developed a method to reconstruct the projected 
density field out to recombination directly 
from the CMB anisotropies \cite{sz98}. The method is based on the 
gravitational lensing effect, which distorts the pattern of 
CMB anisotropies \cite{bern}. 
Although the signal to noise for individual structures from such a 
reconstruction is small,
averaging over independent patches of CMB
reduces the noise and extracts the signal in a statistical 
sense. We were able to show that this allows one 
to extract the power spectrum of density 
perturbations with high accuracy over 
two decades in angular scale \cite{zs98}. 

In the present paper we use the reconstructed projected density field
to cross-correlate it with the CMB itself. If there is a component
to the CMB from the time evolving gravitational
potential then it should correlate with the projected density field. 
Most of the signature comes from large angular scales, so we first 
generalize the method developed in \cite{sz98} to all sky.
Because the small scale CMB anisotropies were generated uniquely 
during recombination, the weighting of density 
perturbations as a function of redshift in the projection is well defined.
Moreover, gravitational lensing effect depends on the dark matter 
distribution in the universe, so no assumption of how light traces 
mass is necessary. 
This avoids the shortcomings of cross-correlations with X-ray and other
tracers of large scale structure mentioned in \cite{ct}.
In addition, the projected density field is sensitive to matter 
distribution 
out to a very high redshift and allows one to test the models where 
the time dependent potential is generating anisotropies at higher 
redshifts, such as the curvature dominated models \cite{kink,marc}. 
Here we compute
the expected signal to noise of future 
CMB missions for cosmological constant 
and curvature dominated models, using both MAP and Planck satellite 
characteristics. Although we limit to these two families of models 
we note that other models, such as those with more general equation 
of state, would also produce a signature that one could look for.

To reconstruct the projected density field we consider the
symmetric tensor of products of temperature derivatives transverse to direction $\hat n$
\cite{sz98}
\begin{equation}
{\cal H}_{ab}={1 \over \sigma_{\cal S}} T_{:a} T_{:b}
\end{equation}
where $T_{:a}$, $T_{:b}$ are covariant derivatives of $T$ with respect to the coordinate basis in 
the tangent space of direction $\hat n$,
here defined with polar coordinates $(\theta, \phi)$ and $g_{ab}$ is the metric 
on the sphere. We defined
$\sigma_{\cal S}$ so that in the absence of lensing the average over CMB
$\langle {\cal H}_{ab} \rangle_{CMB} ={1 \over 2}g_{ab}$. The tensor can be decomposed into the trace 
and traceless component as ${\cal H}_{ab}={1 \over 2}(1-{\cal S})g_{ab}-{\tilde{{\cal H}}_{ab}}$.
From the traceless tensor ${\tilde{{\cal H}}_{ab}}$ one may define two rotationally 
invariant quantities
\begin{equation}
{\cal E}={1 \over 2}\nabla^{-2}{\tilde{{\cal H}}_{ab}}^{:a:b}\;\; \,\,\,{\cal B}={1 \over 2}\nabla^{-2}{\tilde{{\cal H}}_{ab}}^{:a:c}\epsilon^a_c,
\end{equation}
where $\epsilon^a_c$ is the completely antisymmetric (Levi-Civitta) tensor and 
$\nabla^{-2}$ is the inverse Laplacian on the sphere.

In the presence of lensing the average
of ${\cal H}_{ab}$ becomes \cite{sz98}
\begin{equation}
\langle {\cal H}_{ab} \rangle_{CMB} = \left({ 1 \over 2}-\kappa\right)g_{ab}-\gamma_{ab}
\label{avgprime}
\end{equation}
where $\kappa$ and $\gamma_{ab}$ are the convergence and the shear 
components of the symmetric shear tensor $\Phi_{ab}$, defined as the covariant derivative 
of the displacement field on the sphere \cite{stebbins},  
which encodes the information
on the gravitational lensing effect. 
All rotationally invariant quantities can be decomposed on a sphere into 
spherical harmonics,
$X(\hat n)=\sum_{lm} a_{X,lm} Y_{lm}(\hat n)$, 
where $X$ stands for $T$, $\kappa$, $\cal S$, $\cal E$ or $\cal B$. 
From equation [\ref{avgprime}] we find $\langle a_{{\cal S},lm} \rangle = M_{{\cal S}l}a_{\kappa,lm}$,
with $M_{{\cal S}l}=2$.
The multipole moments of the scalar field $\cal E$ average to 
$\langle a_{{\cal E},lm} \rangle = M_{{\cal E}l}a_{\kappa,lm}$, 
where $M_{{\cal E}l}=2(l+2)(l-1)/l(l+1)$ \cite{stebbins},
while the average of 
the pseudo scalar field $\cal B$ identically vanishes in the large scale limit, 
$\langle a_{{\cal B},lm} \rangle =0$, 
because gravitational 
potential from which shear is generated is invariant under the parity 
transformation. Convergence
$\kappa$ can thus be reconstructed 
in two independent ways from $\cal S$ and $\cal E$, 
while the third quantity $\cal B$ serves as 
a check for possible systematics. Note that since convergence is expressed
as a quadratic quantity of $T$ its cross-correlation with $T$ gives a 
non-vanishing 3rd moment. This means it can also be looked for using 
bispectrum, which is a method independently proposed by \cite{sg98}.

Convergence can be written
as a projection of gravitational potential \cite{kaiser,js}
$\kappa= \int_0^{\chi_0} g(\chi,\chi_0)\nabla^2\phi(\chi) d \chi$.
Here $\chi_0$ is the comoving radial coordinate at recombination and
$g$ is the radial window, defined as $g(\chi,\chi_0) = {r(\chi)r(\chi_0-\chi) \over r(\chi_0) }$. It
is a bell shaped curve symmetric around
$\chi/2$ and vanishing at 0 and $\chi_0$.
Here $r(\chi)$ is the comoving angular diameter
distance, defined as $K^{-1/2}\sin K^{1/2}\chi$,
$\chi$, $(-K)^{-1/2}\sinh (-K)^{1/2}\chi$ for $K>0$, $K=0$, $K<0$,
respectively, where $K$ is the curvature.
Curvature
can be expressed using the present density
parameter $\Omega_0$
and the present
Hubble parameter $H_0$ as $K=(\Omega_0-1)H_0^2$.
In general $\Omega_0$ consists both of matter contribution
$\Omega_m$ and cosmological constant term $\Omega_{\Lambda}$.

The angular power spectra are
defined as $C^{XX'}_l= {1 \over 2l+1} \sum_m a_{X,lm}^*a_{X',lm}$. 
Their ensemble 
averages are given by \cite{zsb}
\begin{eqnarray}
C^{XX'}_l=(4\pi)^2 &\int& \beta^2d\beta
P(\beta)\Delta_{Xl}(\beta,\tau_0)\Delta_{X' l}(\beta,\tau_0)
\nonumber \\
\Delta_{X l}(\beta,\tau_0) =&\int_0^{\tau_0}& d\tau \Phi_\beta^{l}(\tau_0-\tau)
S_{X}(\beta,\tau),
\label{cl}
\end{eqnarray}
where $\Phi_\beta^{l}(x)$ are the ultra-spherical Bessel functions and
$P(\beta)$ is the primordial power spectrum.
Equation \ref{cl} only applies to flat and open universes, whereas
for the closed universe
the eigenvalues of the Laplacian are discrete so the
integral over $\beta$
is replaced with a sum over $K^{-{1\over 2}} \beta=3,4,5...$.
The source for temperature anisotropies $S_T$
is a combination of several terms. 
These can be decomposed into terms generated 
during recombination, which consist of Sachs-Wolfe term, Doppler term, 
intrinsic anisotropy term and anisotropic stress term, and late time 
term generated by the time dependent gravitational potential (so-called 
integrated Sachs-Wolfe term or ISW). The latter is only important for low 
multipole moments. 
The full form of $S_T$ can be found in \cite{zsb}. 
The source for convergence is $S_{\kappa}=g \beta^2 \phi$.
It is also useful to define the  
correlation coefficient ${\rm Corr}^{T\kappa}_l=C^{T\kappa}_l/
(C^{TT}_{l}C^{\kappa \kappa}_l)^{1/2}$, 
which is the relevant quantity for the estimation of signal 
to noise. 

\begin{figure}[htbp]
\centerline{\psfig{file=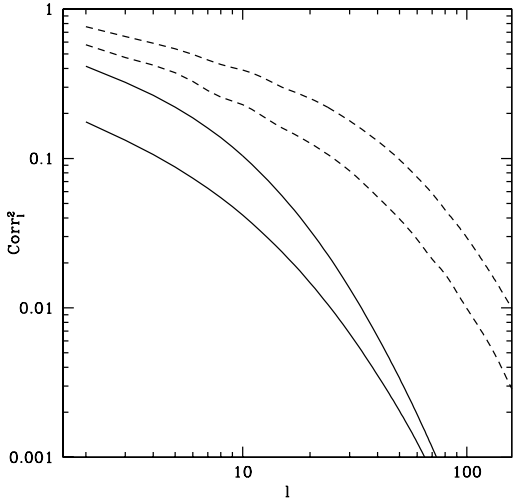,width=2.8in}}
\caption{Square of correlation coefficient $({\rm Corr}^{T\kappa}_l)^2$
as a function of $l$
is shown for open $\Omega_m=0.2$ and $\Omega_m=0.4$ models (upper and 
lower dashed curve) and for cosmological constant 
$\Omega_m=0.2$ and $\Omega_m=0.4$ models (upper and lower solid
curves).}
\label{fig1}
\end{figure}

Using the above expressions one can compute $C^{TT}_l$, $C^{\kappa \kappa}_l$ 
and 
$C^{T\kappa}_l$ for any cosmological model. We performed numerical calculations
using a modified version of CMBFAST code \cite{sz96}. We have verified that 
$C^{\kappa \kappa}_l$ agrees with previous calculations which were done in the 
small scale limit \cite{js}, as well as with the alternative all-sky 
expressions given in
\cite{stebbins}. The results for 
$({\rm Corr}^{T\kappa}_l)^2$ are shown in 
figure \ref{fig1} for a representative set of models. The correlation 
coefficient in a cosmological constant model is substantial 
only for very low $l$
and rapidly drops on smaller scales. In a curvature dominated model 
with the same $\Omega_m$
the correlation coefficient is larger to start with and also drops 
less rapidly with $l$. This indicates that one will be able
to set stronger limits on curvature than on cosmological constant, 
which is confirmed below with a more detailed analysis. The reason 
is that 
in a cosmological constant model the gravitational potential 
changes only at late times ($z\lsim 1$ for reasonable values of 
$\Omega_m$), while in a curvature dominated model potential 
changes also at higher redshift. This leads to two effects. 
First is that in a cosmological constant 
model ISW is comparable to other terms only for the lowest multipoles, 
while in a curvature model ISW dominates up to higher $l$ \cite{marc}. 
Second is that
the window $g$ peaks at relatively 
high redshift $z \sim 3$ and so is able to pick up correlations 
with ISW from open universe better than that from cosmological constant 
universe. We have also calculated the correlation coefficient for
a flat model. It is small for all $l$, $({\rm Corr}^{T\kappa}_l)^2 < 10^{-3}$, 
demonstrating that correlations with 
fluctuations generated at recombination are negligible 
and the cross-correlation is 
sensitive to the time dependent gravitational potential only. 

We now address the question of 
signal detectability with the future CMB missions. We continue to
work in multipole moment space and assume we have all sky expansion, which 
allows us to decouple between different $m$ and $l$ multipole moments. The
generalization to incomplete sky coverage can be approximated by inserting 
appropriate factors of sky coverage fraction $f_{sky}$
in the final expressions.
Given two random fields $T$ and $\cal W$ (where $\cal W$ stands for $\cal S$ or $\cal E$) 
we want to develop a test 
that maximizes the signal in the presence of correlations against the 
null hypothesis that there are no correlations. The term that quantifies
the correlations is the product between the two fields $X^{\cal W}_{lm}=T_{lm}^*
{\cal W}_{lm}$ (here and below average with the complex conjugate 
$(T_{lm}^*{\cal W}_{lm}+T_{lm}{\cal W}_{lm}^*)/2$ is implied). 
Its expectation value under the null hypothesis of pure noise 
is $\langle X^{\cal W}_{lm} \rangle_0=0$, because the function entering this 
expression is a three-point function of $T$, which vanishes both for 
intrinsic fluctuations and for detector noise under the Gaussian assumption. 
The alternative hypothesis is that of signal which gives
$\langle X^{\cal W}_{lm} \rangle_1 = \langle T_{lm}^* {\cal W}_{lm} \rangle= M_{{\cal W}l}C^{T \kappa}_l$.
The variance under the null hypothesis is
\begin{eqnarray}
&\langle& X^{\cal W}_{lm}X^{\cal W'}_{lm} \rangle_0-
\langle X^{\cal W}_{lm} \rangle_0\langle X^{\cal W'}_{lm} \rangle_0= \nonumber \\
&(&C^{TT}_{l}+N^{TT}_l)
(M_{{\cal W}l}M_{{\cal W'}l}C^{\kappa \kappa}_{l}+N^{\cal W W'}_l),
\end{eqnarray}
where $N^{TT}_l$ and $N^{\cal WW'}_l$ are the noise power spectra for 
CMB anisotropies, $\cal S$, $\cal E$ or their cross-term, respectively. 

\begin{figure}[htbp]
\centerline{\psfig{file=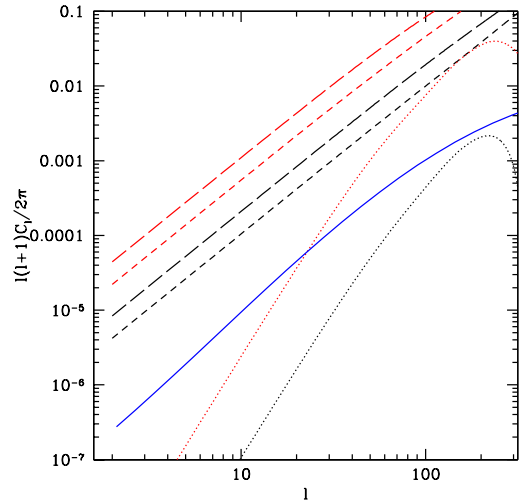,width=2.8in}}
\caption{Power spectra for noise $N^{\cal S \cal S}_l$ (long dashed), 
$N^{\cal E\cal E}_l$ (short dashed), $N^{\cal E \cal S}_l$ (dotted)
both for Planck (lower curves) and MAP (upper curves) for cosmological 
constant model with $\Omega_m=0.3$. Also shown is power 
spectrum of convergence $ 4C^{\kappa \kappa}_l$ (solid) for the same model, 
normalized to $\sigma_8=1$.}
\label{fig2}
\end{figure}

Both $\cal S$ and $\cal E$ contribute information. If they are uncorrelated then 
the information contents can be added independently, otherwise the
covariance matrix ${\rm Cov}(X^{\cal W}_{lm}X^{\cal W'}_{lm})$ has to be diagonalized first.
The CMB term $C^{TT}_{l}+N^{TT}_l$ is the same for all matrix elements and can be computed 
using MAP and Planck noise characteristics. For these
CMB missions detector noise on large scales will be negligible, hence 
$N^{TT}_l \ll C^{TT}_{l}$.
The dominant source of noise in $\cal S$ or $\cal E$ on large scales are the CMB anisotropies.
The noise terms involve integrals over the CMB power spectrum and 
can be computed numerically using the expressions
given in \cite{sz98}. The results are shown in figure \ref{fig2} both for MAP and Planck. 
They show that on large scales the CMB noise
power spectrum has approximately white noise shape, $N^{\cal W \cal W}_l \sim \rm{ const}$. 
At low $l$
$N^{\cal E \cal E}_l \sim N^{\cal S\cal S}_l/2 \gg N^{\cal E \cal S}_l, C^{\kappa \kappa}_l$.
Therefore, noise dominates over the signal and the latter can only be extracted 
in a statistical sense by averaging over multipole moments.
Because the off-diagonal term $\langle {\cal E}_{lm}^* {\cal S}_{lm} \rangle_0$
is much smaller than the two diagonal terms the covariance matrix  
is nearly diagonal and the information from
$\cal S$ and $\cal E$ can be added independently, with 
$\cal E$ contributing twice the amount of information than $\cal S$. Note also that
Planck has a factor of 5 better sensitivity than MAP.

We now want to combine the signal to noise from different multipole 
moments 
to maximize the overall signal to noise. To do this we 
add up the products weighted with some yet to be determined weights $\alpha_l$,
$X=\sum_{m,l} \alpha_l X_{lm}$. Since the moments are uncorrelated the 
expectation value and variance are 
\begin{eqnarray}
\langle X\rangle_1 &=& \sum_{l} (2l+1)\alpha_l M_{{\cal W},l}C^{T\kappa}_{l} \nonumber \\
\langle X^2\rangle_0 &=& 
\sum_{l} (2l+1)\alpha_l^2 (C^{TT}_l+N^{TT}_l)(M_{{\cal W}l}^2C^{\kappa \kappa}_l+N^{\cal WW}_l),
\end{eqnarray}
while the null hypothesis mean remains $\langle X\rangle_0=0$.
We want to maximize $S/N=(\langle X\rangle_1-\langle X\rangle_0)/(\langle X^2\rangle_0)^{1/2}$
with respect to $\alpha_l$. Taking derivatives with respect to $\alpha_l$
and setting the expression to 0 we find $\alpha_l=C^{T\kappa}_{l}/
(C^{TT}_{l}+N^{TT}_l)(M_{{\cal W}l}^2C^{\kappa \kappa}_l+N^{\cal W \cal W}_l)$. The overall signal to noise
is, combining the information from $\cal S$ and $\cal E$
\begin{equation}
{ S \over N}= \left[ f_{sky}\sum_l\sum_{\cal W=\cal E, \cal S}
{ (2l+1)({\rm Corr}^{T \kappa}_l)^2 \over (1+{N^{TT}_l \over C^{TT}_{l}})(
1+{N^{\cal W \cal W}_l \over M_{{\cal W}l}^2C^{\kappa \kappa}_l})}\right]^{1/2},
\label{sn}
\end{equation}
where we inserted $f_{sky}$ to account for the fact that the 
number of multipoles will be smaller if only some fraction of the sky will 
be measured. 
We have expressed the signal in terms of number of standard 
deviations above the noise. To express it in terms of confidence limits a more 
detailed analysis with Monte Carlo simulations is needed \cite{sz98}, but 
in the limit where many multipole moments contribute to the signal 
the usual Gaussian confidence limits as a function
of number of standard deviations apply.
The above expression shows that if correlation is unity and noise 
is negligible then each 
multipole moment contributes one degree of freedom and the signal to 
noise is as expected $N_{\rm dof}^{1/2}$, where $N_{\rm dof}$ is the number 
of degrees of freedom. Decorrelation and/or noise decrease the effective number
of degrees of freedom. 

Using above expressions we find $S/N=8$ for $\Omega_m=0.4$ open model
and $S/N=13$ for $\Omega_m=0.2$ open model, both for Planck noise and 
beam properties using $f_{sky}=0.7$. 
Corresponding numbers for MAP are 3.5 and 7.
Both MAP and Planck will thus be able to usefully constrain open models with 
$\Omega_m<0.4$, 
which spans the range of currently favored values of $\Omega_m$. For 
cosmological constant model the numbers are somewhat lower, Planck 
gives $S/N=3$ and 6 for $\Omega_m=0.4$ and $\Omega_m=0.2$, 
respectively, while corresponding MAP numbers are 1 and 2. A positive 
detection in these models 
can therefore only be obtained with Planck, unless $\Omega_m$ is very low. One
can use the absence or presence of cross-correlation to put constraints on 
the models. 
Any detection of the signal with MAP will for example be more 
easily explained in terms of curvature models than with 
cosmological constant models, while absence of the signal in Planck will certainly
rule out all curvature models of interest, as well as put strong constraints on 
cosmological constant models. Within the context of more specific models, such as the
family of CDM models, one can 
use the cross-correlation to break the degeneracies present when only 
the CMB power spectrum constraints are used. The well-known degeneracy between curvature 
and cosmological constant can for example be broken using this cross-correlation. A more 
detailed analysis which includes temperature, polarization and convergence
information will be presented elsewhere \cite{hsz98}. Note that the theoretical 
limit for signal to noise can be obtained by assuming $\kappa$ is perfectly known and
is given by $S/N=\sum_l (2l+1)({\rm Corr}^{T \kappa}_l)^2$. This gives 
$S/N$ about a factor of 2 higher than our results for Planck above.

Finally, we should mention 
that signal should be consistent with the null hypothesis (pure noise) 
for the $\cal B$ field in the large scale limit. 
Any evidence against that would be a sign of a systematic effect present 
in the data. 
This test provides a useful overall check of the method.
Another useful test would be cross-correlating $\cal E$ and $\cal S$
with the polarization CMB map. Since ISW does not contribute to the latter
the result should again be consistent with zero. 
The straightforward interpretation and many consistency checks 
make the here proposed method one of the most 
promising ways to determine cosmological parameters and 
should provide further incentive for high sensitivity all-sky CMB experiments.

U.S. and M.Z. would like to thank Observatoire de Strasbourg and MPA,
Garching, respectively, for
hospitality during the visits.
M.Z. is supported by NASA through Hubble Fellowship grant
HF-01116.01-98A from STScI,
operated by AURA, Inc. under NASA contract NAS5-26555.

\vfil\eject
 
\end{document}